\documentclass[prb,twocolumn,showpacs,amsmath,amssymb,superscriptaddress]{revtex4}
\usepackage{graphicx}

\begin{document}

\title{Disentangling thermal and non-thermal excited states in a charge-transfer insulator by time-and-frequency resolved pump-probe spectroscopy}

\author{Claudio Giannetti}
\affiliation{Department of Physics, Universit$\grave{a}$ Cattolica del Sacro Cuore, Brescia I-25121, Italy.}
\author{Goran Zgrablic}
\affiliation{Sincrotrone Trieste S.C.p.A., Basovizza I-34012, Italy.} 
\author{Cristina Consani}
\altaffiliation[Present address: ]{\'Ecole Polytechnique F\'ed\'erale de Lausanne (EPFL), Laboratoire de Spectroscopie Ultrarapide, CH-1015 Lausanne, Switzerland.}
\affiliation{Department of Physics, Universit$\grave{a}$ degli Studi di Trieste, Trieste I-34127, Italy.}
\author{Alberto Crepaldi}
\altaffiliation[Present address: ]{\'Ecole Polytechnique F\'ed\'erale de Lausanne (EPFL), Institut de Physique des Nanostructures, CH-1015 Lausanne, Switzerland.}
\affiliation{Department of Physics, Universit$\grave{a}$ degli Studi di Trieste, Trieste I-34127, Italy.}
\author{Damiano Nardi}
\affiliation{Department of Physics, Universit$\grave{a}$ Cattolica del Sacro Cuore, Brescia I-25121, Italy.}
\affiliation{Department of Physics, Universit$\grave{a}$ degli Studi di Milano, I-20122, Italy.}
\author{Gabriele Ferrini}
\affiliation{Department of Physics, Universit$\grave{a}$ Cattolica del Sacro Cuore, Brescia I-25121, Italy.}
\author{G. Dhalenne}
\affiliation{Laboratoire de Physico-Chimie de l'Etat Solide, ICMMO, CNRS, UMR8182, Universit\'e Paris-Sud, Bat 414, 91405 Orsay, France.}
\author{A. Revcolevschi}
\affiliation{Laboratoire de Physico-Chimie de l'Etat Solide, ICMMO, CNRS, UMR8182, Universit\'e Paris-Sud, Bat 414, 91405 Orsay, France.}
\author{Fulvio Parmigiani}
\affiliation{Sincrotrone Trieste S.C.p.A., Basovizza I-34012, Italy.} 
\affiliation{Department of Physics, Universit$\grave{a}$ degli Studi di Trieste, Trieste I-34127, Italy.}

\date{\today}

\begin{abstract}
Time-and-frequency resolved pump-probe optical spectroscopy is used to investigate the effect of the impulsive injection of delocalized excitations through a charge-transfer process in insulating CuGeO$_3$. A large broadening of the charge-transfer edge is observed on the sub-ps timescale. The modification of this spectral feature can not be attributed to the local increase of the effective temperature, as a consequence of the energy absorbed by the pump pulse. The measured modifications of the optical properties of the system are consistent with the creation of a non-thermal state, metastable on the ps timescale, after the pump-induced impulsive modification of the electron interactions.
\end{abstract}

\maketitle
\section{Introduction}
\label{Sec1}
In the last years, the investigation of the dynamics of strongly correlated systems (SCS) far from equilibrium emerged as an important but difficult and intriguing problem.\cite{Freericks:2003} The fundamental questions are related to the possibility of impulsively modifying and, possibly, controlling the many-body electronic local interactions. 

However, to approach this problem some fundamental questions must be considered. For example, how are described the dynamics of the electronic band structure, as the hamiltonian is modified under non-equilibrium conditions? Does the system relax to a quasi-thermal state corresponding to an increase of the effective local temperature after energy absorption or is the interaction so deeply modified that new non-thermal states can emerge?

Ongoing theoretical works, aimed to investigate the problem of non-equilibrium physics in quantum and correlated systems, are based on real-time diagrammatic Monte-Carlo,\cite{Schiro:2009} adaptive time-dependent density-matrix renormalization group \cite{Manmana:2009} and non-equilibrium Dynamical Mean-Field Theory (DMFT) \cite{Eckstein:2008,Eckstein:2008b} techniques. In particular, the application of DMFT to the Falicov-Kimball model \cite{Eckstein:2008b} evidenced that, after an impulsive variation of the local electronic interaction (interaction quench), the system relaxes to a non-thermal steady state, i.e. a state not corresponding to the increase of temperature $\delta \mathrm{T}$ related to the increase of energy absorbed by the system. The creation of a non-thermal state is possible even in the case of an infinitesimal interaction quench, i.e. when the impulsive change of the local interaction is smaller than the critical interaction $U_c$, i.e. the energy necessary for the occurrence of a metal-insulator transition.\cite{Eckstein:2008b}

Optical time-resolved pump-probe experiments are a unique tool to investigate this physics, being able to directly follow the sub-ps relaxation dynamics of a system, after excitation with a fs pump pulse. Different configurations, such as time-resolved optical and photoemission spectroscopies have been recently employed to investigate the dynamics of photo-induced Mott metal-insulator transitions.\cite{Ogasawara:2000,Iwai:2003,Chollet:2005,Perfetti:2006,Okamoto:2007,Kubler:2007} However, for investigating the non-equilibrium dynamics of a system, in the limit of small interaction quenches, it is mandatory to develop an experimental technique capable to detect small variations of the electronic properties in a wide frequency spectral range and time domain, and to elaborate new interpretative models. 

Here we report on a time-resolved optical spectroscopy with simultaneous high time and frequency resolution applied to study the non-equilibrium electronic structure of CuGeO$_3$ when a charge-transfer process is photo-induced by a suitable light pulse in the fs time domain. The non-equilibrium optical response is investigated by ultra-short white light pulses, in a spectral range extending from 400 nm to 800 nm. 

CuGeO$_3$ is characterized by the "robustness" of the insulating phase, i.e. the absence of transitions to metallic states at high temperatures, and by 1-D non-interacting chains of edge-sharing Cu$-$O$_4$ plaquettes. The fine frequency resolution is exploited to observe in real-time the non-equilibrium dielectric function and to follow the line profile variation of the spectral features, disentangling the thermal and non-thermal effects. In particular, we show that the photo-injection of delocalized excitations within the Cu$-$O$_4$ plaquettes, obtained by the impulsive excitation of the charge transfer transition from the Cu-$3d$ localized states into the O-$2p$ delocalized bands, strongly perturbs the Cu-$3d$$-$O-$2p$ transfer integral, modifying the hamiltonian and leaving the system in an excited state which is not compatible with the thermal state expected considering the local increase of temperature related to the energy absorbed by the pump pulse.

\section{Thermal vs non-thermal excited states in a charge-transfer insulating cuprate}
\label{Sec2}
Copper oxide based compounds (cuprates), are one of the most important SCS family, because of the high-T$_C$ superconducting phase shown by some doped systems. In general, copper oxides exhibit a variety of copper coordinations ranging from square-planar to octahedral and distorted octahedral. In these compounds the Cu-$3d$ orbitals contain an odd number of electrons and a metallic ground state should be expected. In contrast with the predictions of the one-electron band-theory, the Coulomb energy related to the on-site electron-electron interactions overwhelms the energy width of the conduction band related to the translational symmetry, accounting for the insulating ground-state.\cite{Eskes:1990,Atzkern:2001} Cluster models (CM) have been developed to calculate, for a simple Cu$_x$O$_y$ cluster,\cite{Eskes:1990} the eigenvalues of the hamiltonian $H$=$\sum_lE_d(l)d^{\dag}_ld_l$+$\sum_lE_p(l)p^{\dag}_lp_l$+$\sum_lT_{pd}(l)(d^{\dag}_lp_l+p^{\dag}_ld_l)$+$H_U$, $d$ and $p$ being operators creating a hole in the Cu-$3d$ and O-$2p$ orbitals, indexed by the spin and orbital quantum number $l$ ($m,n,m',n'$) and $T_{pd}$ the Cu-$3d$$-$O-$2p$ transfer integral. Electron correlations are mimicked through the interaction hamiltonian $H_U$=$\sum^{n,n'}_{m,m'}U(n$,$n'$,$m$,$m')d^{\dag}_md_{m'}d^{\dag}_nd_{n'}$, being $U$ the repulsive local interaction. The large value of the $U$ term is responsible for the insulating gap of the undoped cuprates. Cluster calculations qualitatively account for the charge-transfer nature of the insulating gap in undoped cuprates. 

When a charge transfer insulator, at fixed temperature T, absorbs a short and intense laser pulse, the nature of the excited state can follow two different paths. i) Quasi-thermal excited state. In this case, the system is left in a thermal state described by an effective temperature $\mathrm{T}+\delta \mathrm{T}$, where the temperature increase corresponds to the increase of the internal energy of the system delivered by the pump pulse. The action of the pump pulse merely corresponds to a variation of the electronic distribution. In this case the relaxation dynamics can be efficiently described by the two-temperatures model.\cite{Anisimov:1974} ii) Non-thermal excited state. In this scenario, the excitation perturbs the $T_{pd}$ and $U$ terms of the hamiltonian to an extent that the ground state impulsively changes and the dynamics of the physical properties is characterized by the relaxation toward the new ground state. This new state does not correspond to a quasi-thermal state labeled by an effective temperature $\mathrm{T}+\delta \mathrm{T}$. 

The most striking evidence of non-thermal states is the possibility to photoinduce metal-insulator phase transitions in a Mott insulator.\cite{Perfetti:2006} However, less efforts have been devoted to investigate the dynamics of charge-transfer insulating cuprates, which do not exhibit any metal phase even when $T_{pd}$ and $U$ are strongly modified by the fs pump pulses. In this case, the collapse of the insulating gap during the non-thermal insulator-to-metal phase transition is replaced by a weaker variation of the electronic properties of the insulating phase. For this reason it is mandatory to develop an experimental set-up with high time and frequency resolution able to follow the variation of the spectral features typical of a charge-transfer insulator on the sub-ps timescale.

\begin{figure}
\includegraphics[keepaspectratio, bb= 0 0 480 390, clip,width=0.48\textwidth]{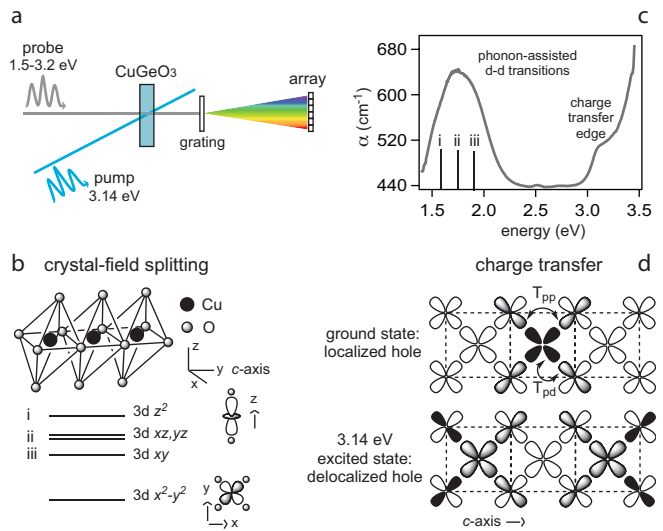}
\caption{\label{fig1} (Color online) (a) Schematics of the blue pump-SWL probe experiment. (b) The sketch of the edge sharing Cu$-$O$_6$ octahedra is reported. Splitting of the level induced by the local crystal field is shown. (c) Absorption coefficient in the visible spectrum with light polarization along the c-axis, taken from Ref. \onlinecite{Pagliara:2002}. (d) Charge distribution within the Cu$-$O$_4$ plaquettes in the ground state and delocalized excited state, as calculated through cluster models on Cu$_3$O$_8$ clusters. High, medium and low hole density is indicated by the black, gray and white colors.}
\end{figure}

\section{Experimental}
\label{Sec3}

The time-and-frequency resolution is obtained by focusing 60 fs-790 nm light pulses on a CaF$_2$ platelet for producing supercontinuum white-light (SWL) at 1 kHz repetition rate. The SWL is focused on the sample in spatial and temporal coincidence with the 400 nm or 800 nm pump pulse and dispersed on a 512 pixel NMOS array, after the interaction (see Fig. \ref{fig1}a). A complete scan of the variation of the transmissivity $\mathcal{T}$ is triggered by the pump pulse and completed before the following laser pulse. At the beginning, both the pump and probe polarizations are directed along the \textit{c}-axis. The relative variation of the transmissivity $\mathcal{T}$ is directly related to the pump-induced change of the absorption coefficient $\alpha$ through $\delta \mathcal{T}(t)/\mathcal{T}$=$-L\delta \alpha (t)$, being the sample thickness $L$=20 $\mu$m. The temporal profile of SWL is characterized by means of optical Kerr effect in water. CuGeO$_3$ single crystals were grown from the melt by a floating zone technique.\cite{Revcolevschi:1993}\\ 

\begin{figure}
\includegraphics[keepaspectratio, bb= 0 180 580 660, clip,width=0.48\textwidth]{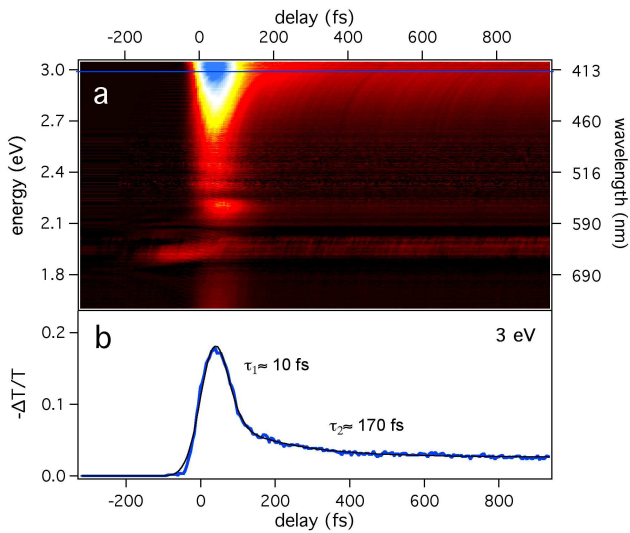}
\caption{\label{fig2} (Color online)(a) Time-and-frequency transmissivity relative variation of CuGeO$_3$ for 3.1 eV excitation energy. Both the pump and probe beams are polarized along the $c$-axis. The absorbed pump fluence is $\Phi_{pump}$=20 mJ/cm$^2$. (b) Time trace at fixed photon energy (3 eV). The solid line is the fit to the data using a double exponential decay convoluted with a gaussian curve.}
\end{figure}
\section{C\lowercase{u}G\lowercase{e}O$_3$ crystal structure and electronic properties}
The relevant unit of the system is a planar Cu$-$O$_4$ plaquette. In particular, while in high-temperature superconductors the Cu$-$O$_4$ units share their corners, in CuGeO$_3$
the Cu plaquettes are edge-sharing coordinated along the \textit{c}-axis (see Fig. \ref{fig1}b), the Cu-$3d$ and O-$2p$ orbitals forming a $\sim$98$^{\circ}$ angle. At room temperature CuGeO$_3$ belongs to the orthorhombic group D$_{2h}$. The unit cell contains two edge-sharing strongly deformed Cu$-$O$_6$ octahedra, with distances Cu-O$_{apex}$=2.77 $\AA$ and Cu$-$O$_{inplane}$=1.94 $\AA$.\cite{Vollenkle:1967}\\

\begin{figure*}
\includegraphics[keepaspectratio, bb= 30 10 750 530, clip,width=0.92\textwidth]{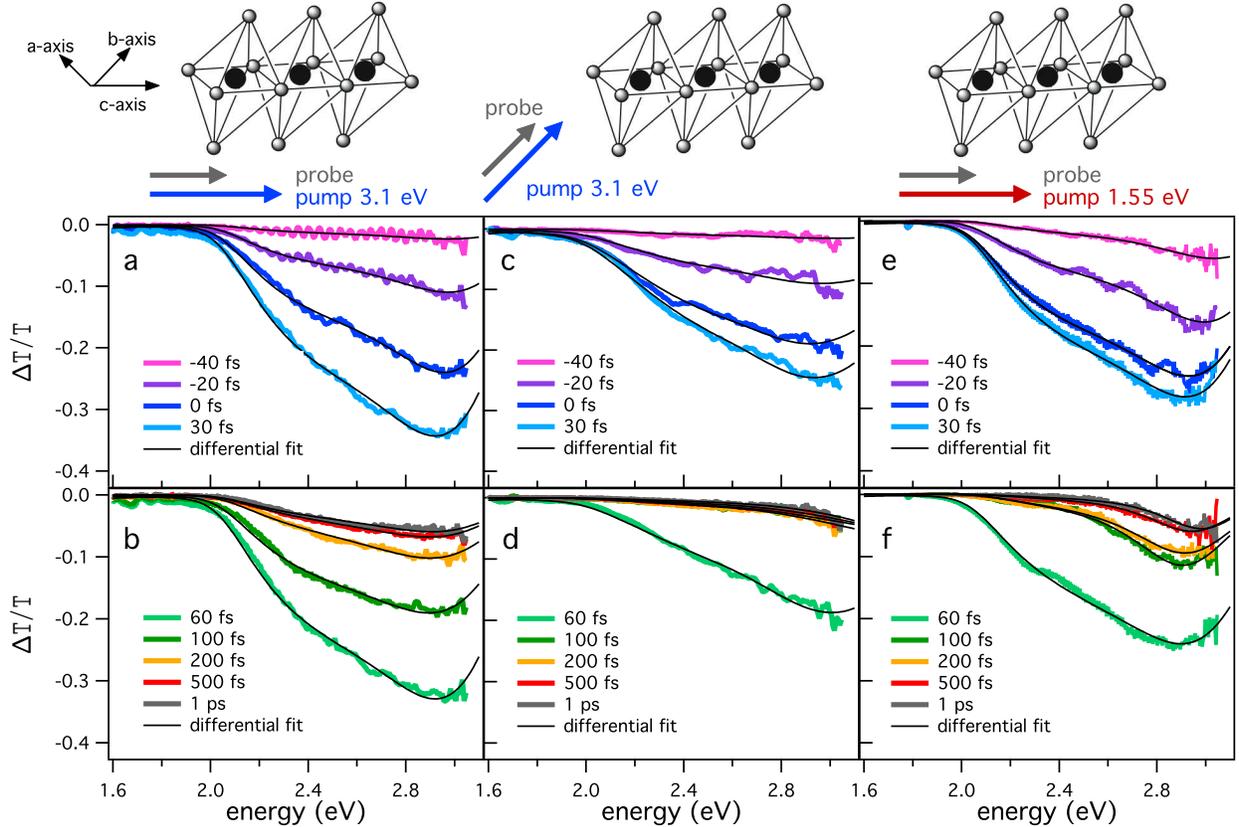}
\caption{\label{fig3} (Color online) Slices of the two-dimensional $\Delta$$\mathcal{T}$/$\mathcal{T}$($\omega$,t) spectrum at different delay times and different pump polarizations and wavelengths. The solid lines are the fit to the data of the differential dielectric function.}
\end{figure*}
Because of its peculiar crystal and electronic structure, CuGeO$_3$ is a quite suitable simple system to study the relaxation dynamics of a charge-transfer insulating cuprate upon the impulsive injection of excitations.
Fig. \ref{fig1}c reports the peculiar features of the optical spectrum, when the incident light is polarized along the c-axis. The absorption band at 1.5-2.1 eV is related to phonon-assisted transitions between Cu-$3d$ levels split as a consequence of the crystal field experimented by the Cu atoms.\cite{Bassi:1996,Pagliara:2002} In particular three oscillators can be identified at photon energies of $\sim$1.6 eV, 1.75 eV and 1.9 eV, attributed to transitions from the 3$d_{x^2-y^2}$ hole ground state to 3$d_{xy}$, 3$d_{xz,yz}$ and 3$d_{z^2}$ levels \cite{Broer:2000} (see Fig. \ref{fig1}b). At photon energies larger than $\sim$3 eV, the edge in the absorption coefficient (see Fig. \ref{fig1}c) is attributed to the charge transfer process from the ground state (hole in the $Cu$ atom) to an excited state (hole in the $O$ atoms).\cite{Bassi:1996} Considering the hybridization of the O-$2p$ orbitals with both the Cu-$3d$ and the nearest-neighbor O-$2p$ orbitals, the effective charge transfer gap for the $x^2$-$y^2$ symmetry state is $\sqrt{(\Delta_{pd}-T_{pp})^2+4T_{pd}^2}$, being $\Delta_{pd}$ the energy of an hole in the O-$2p$ orbital, $T_{pd}$ the overlap integral between O-$2p$ and Cu-$3d$ orbitals and $T_{pp}$ the overlap integral between O-$2p$$-$O-$2p$ orbitals, including both $\sigma$ and $\pi$ bondings. Note that $T_{pp}$$\neq$0 (and hopping of the hole to adjacent plaquettes is possible) only if the Cu$-$O$-$Cu bonding angle is different from 90$^{\circ}$. The shoulder in the absorption coefficient at 3.1-3-2 eV (see Fig. \ref{fig1}c) is attributed to a superposition of plasmons related to transitions from the ground state, being 72$\%$ of the hole density localized on Cu atoms, into a delocalized final state with a high occupation probability of the neighbor plaquettes,\cite{Atzkern:2001,Pagliara:2002} as shown in Fig. \ref{fig1}d. The final state corresponds to the formation of a spin-singlet state on the neighboring plaquette, known as Zhang-Rice singlet.\cite{Zhang:1988} As a consequence of dipole selection rules, this excitonic transition can be induced only when the polarization of the light is parallel to the c-axis. If the polarization is perpendicular to the c-axis (b-axis), only the charge transfer edge survives, corresponding to injection of localized holes in the O-$2p$ orbitals, whereas the excitonic shoulder does not contribute.

The shape of the charge transfer edge (see Fig. \ref{fig1}c), in both c- and b-axis polarizations, follows the Urbach's rule:\cite{Urbach:1953}
\begin{equation}
\label{urbach}
\alpha=\alpha_0e^{\left(\sigma\frac{E-E_0}{k\mathrm{T}}\right)}
\end{equation}
$\alpha_0$ and $E_0$ being temperature independent parameters and $\sigma$ the temperature dependent slope of $k\mathrm{T}$ln$(\alpha/\alpha_0)$, given by $\sigma$=$\sigma_0(2k\mathrm{T}/E_P)$tanh$(E_P/2k\mathrm{T})$, where $E_P$ is the characteristic phonon energy. Best fit values for $E_0$ were determined to be 3.46 eV and 3.67 eV for light polarization parallel to the c- and b-axis, respectively.\cite{Bassi:1996} The Urbach shape of the absorption edge is a universal property of excitonic absorption edges related to exciton scattering by optical phonons.\cite{SchmittRink:1981,Dow:1972} The values $E_P$=53 meV and $E_P$=56 meV extracted from the fit of expression (\ref{urbach}) to the CuGeO$_3$ absorption coefficients with light polarization along the c-axis and b-axis, respectively, match with the energy of a $B_{2g}$
bond-bending mode of oxygen atoms at 427 cm$^{-1}$ (see Ref. \onlinecite{Popovic:1995}), confirming the coupling of the electronic charge transfer process to the position of the oxygen atoms within the unit cell.

\section{Results}
\label{Sec4}
Impulsive injection of delocalized excitations is obtained by optical pumping with coherent light pulses at 3.14 eV and polarized along the c-axis. In Fig. \ref{fig2}a the relative transmissivity variation $\Delta$$\mathcal{T}$/$\mathcal{T}$($\omega$,t) of CuGeO$_3$, upon excitation by 20 mJ/cm$^2$ pump fluence, is reported for time delays between $\sim$-250 fs and $\sim$900 fs and in the spectral range between $\sim$770 nm (1.6 eV) and $\sim$410 nm (3 eV), covered by the supercontinuum probe pulse. The two-dimensional trace directly contains information on the variation of the imaginary part of dielectric function during the photoexcitation of the delocalized holes in the system. The insulating charge-transfer gap at h$\nu$$<$3.4 eV appears to be impulsively quenched, resulting in a $\Delta$$\mathcal{T}$/$\mathcal{T}$($\omega$) variation decreasing from the blue to the near-IR region (1.6 eV). $\Delta$$\mathcal{T}$/$\mathcal{T}$($\omega$,t) scales linearly, as the pump fluence is increased, up to an energy where permanent damages are induced in the sample. In Fig. \ref{fig2}b we report the time-trace of $\Delta$$\mathcal{T}$/$\mathcal{T}$($\omega$) at 3 eV-photon energy, evidencing the dynamics of the transmissivity variation. Three different timescales are observed. A fast one ($\tau_1\sim10$ fs), smaller than the pulse timewidth,\cite{note0} a slower one ($\tau_2\sim170$ fs) responsible for the relaxation on the sub-ps timescale and a very slow one that extends to several hundreds of ps. While the value of $\tau_1$ assesses the electronic nature of this fast process, the slower dynamics, revealed by $\tau_2$, contains the information on the energy relaxation process (cooling), through the electron-phonon coupling, for both the photo-injected holes in the O-$2p$ orbitals and the electrons in the Cu-$3d$ correlated bands, eventually leading to the formation of a new metastable electronic and/or structural phase. The two fast dynamics are evident reporting the time-resolved spectra of $\Delta$$\mathcal{T}$/$\mathcal{T}$($\omega$,t) at different delay times (see Fig. \ref{fig3}a). The transmissivity is impulsively quenched within the pulse duration ($\sim$60 fs), monotonically decreasing ingoing from the near-IR ($\sim$1.6 eV) to the blue ($\sim$3 eV), with a plateau around 3 eV and a kink at about 2.3 eV. After the excitation with the pump pulse, $\Delta$$\mathcal{T}$/$\mathcal{T}$($\omega$,t) relaxes toward the equilibrium, maintaining, for several hundreds of ps, a broad transmissivity variation in the 2-3 eV range (see the spectrum at 1 ps in Fig. \ref{fig3}b). \\

The physics emerging from these observations is summarized in the following. The pump pulses at 3.14 eV impulsively inject a high density of excitations in the Cu$-$O$_4$ plaquettes. The absorbed fluence can be reported in terms of the ratio between the absorbed density of photons $n_{ph}$ and the density of copper ions $n_{Cu}$$\simeq$17$\cdot$10$^{21}$ cm$^{-3}$. In particular, we obtain $n_{ph}/n_{Cu}$$\approx$$\Phi_{pump}\alpha /n_{Cu}$$\simeq$0.1-0.2$\%$, being $\alpha$=500 cm$^{-1}$. Considering the rapid delocalization of each hole within about 5 neighboring plaquettes,\cite{Atzkern:2001} it is possible to conclude that, during the pulse duration, $\sim$0.5-1$\%$ of the Cu$-$O$_6$ units are driven to the excited state where the correlated 3$d$ bands are locally filled by an electron and a hole is shared among the O-$2p$ orbitals and the Cu-3$d$ levels of the neighboring plaquettes (see Fig. \ref{fig1}d), possibly altering both the interaction $U$ and the transfer integral $T_{pd}$, and leaving the system in a metastable non-thermal phase. On the ps timescale, the non-equilibrium holes in the plaquettes relax following two possible pathways: i) they can directly couple to the lattice, increasing its effective temperature; ii) they can readily interact with the Cu-3$d$ equilibrium holes, creating a non-equilibrium population within the Cu 3$d$ bands. To disentangle these effects and to clarify the role of the delocalized excitons we repeated the $\Delta$$\mathcal{T}$/$\mathcal{T}$($\omega$,t) measurements, maintaining constant the absorbed pump fluence, at: i) 3.14 eV pump photon energy with polarization along the b-axis. In this case, the excitonic delocalized transitions are strongly quenched and the relaxation dynamics of localized holes in the O-$2p$ orbitals can be accessed; ii) 1.55 eV pump photon energy with polarization along the c-axis. Under these conditions, the pump energy is exclusively absorbed by phonon-assisted Cu 3$d$-3$d$ intraband transitions and the dynamics is expected to be dominated by the cooling of the non-equilibrium distribution of excitations, within the Cu 3$d$ bands, through electron phonon scattering. 

In Fig. \ref{fig3}c and \ref{fig3}d we report the fast and slow dynamics of $\Delta$$\mathcal{T}$/$\mathcal{T}$($\omega$,t) with both the pump and probe polarization parallel to the b-axis. In this case, the formation of delocalized excitons within the Cu$-$O$_4$ plaquette chains is quenched by the dipole selection rules. The fast dynamics is similar to the previous case, whereas the $\Delta$$\mathcal{T}$/$\mathcal{T}$($\omega$,t) variation, on the ps timescale, is drastically quenched and, $\sim$100 fs after the pump pulse, the spectrum in the 2.8-3 eV energy range shows minor variations. After about 1 ps, the time-resolved spectra do not change their shape sensitively, remaining constant for several hundreds of ps.

In Fig. \ref{fig3}e and \ref{fig3}f we report the fast and slow dynamics of $\Delta$$\mathcal{T}$/$\mathcal{T}$($\omega$,t) by pumping the system using pulses at 1.55 eV linearly polarized along the c-axis of the crystal. The main differences from the previous cases are detected for spectra taken for 100 fs$<$t$<$1 ps (Fig. \ref{fig3}f), where $\Delta$$\mathcal{T}$/$\mathcal{T}$($\omega$,t) exhibits a maximum around 2.9 eV and a kink at about 2.3 eV. After about 1 ps from the pump excitation, a large variation of the transmissivity, mostly in the 2.6-3 eV energy region, is measured, remaining constant for several hundreds of ps.

\begin{figure}
\includegraphics[keepaspectratio, bb= 10 120 1050 730, clip,width=0.92\textwidth]{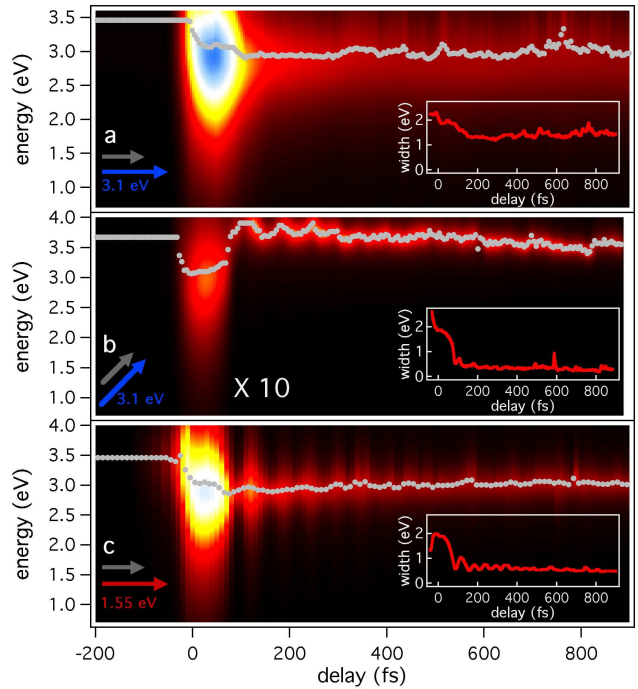}
\caption{\label{fig4} (Color online) (a,b,c) Temporal dynamics of Im$\epsilon_{L}$ at different pump polarizations and wavelengths, as obtained from the differential fitting procedure. The color scales have been normalized to the maximum variation detected. The color scale in Figure \ref{fig4}b has been multiplied by a factor 10. The gray dots indicate the resonance frequencies $\omega_{0}$. The insets display the temporal dependence of the Lorentzian linewidths $\gamma$.}
\end{figure}

\section{Discussion}
\label{Sec5}
To perform a quantitative study of the $\Delta$$\mathcal{T}$/$\mathcal{T}$($\omega$,t) measurements, a differential dielectric function model has been fitted to the data, i.e. the variation of $\epsilon$($\omega$) has been calculated as $\delta$$\epsilon$($\omega$)=$\epsilon_{L}$($\omega$)-$\epsilon_{0}$($\omega$), $\epsilon_{0}$($\omega$) being the dielectric function of CuGeO$_3$ in the ground state.\cite{note1} Notably, the best fit to the data has been obtained by employing, for all the time delays and all the pump polarizations and wavelengths, a single Lorentz oscillator $\epsilon_{L}$($\omega$)=$\omega^2_{P}$/($\omega^2_{0}$-$\omega^2$-i$\gamma$$\omega$), $\omega_{P}$ and $\omega_{0}$ being the plasma and resonance frequencies and $\gamma$ the linewidth. In Fig. \ref{fig3} the best differential fits are superimposed to the data. The main features of the measured $\Delta$$\mathcal{T}$/$\mathcal{T}$($\omega$,t) are reproduced by a Lorentz oscillator centered in the blue region of the spectrum. This result suggests that the physics behind the measured variation of the spectrally-resolved transmissivity is essentially related to the variation of the charge-transfer edge.

Fig. \ref{fig4}a reports the imaginary part of $\epsilon_{L}$($\omega$) during the relaxation process when the 3.1 eV-pump is polarized along the c-axis. The resonance frequency, starting from the equilibrium value $\omega_{0}$=3.46 eV, decreases to 2.9 eV within $\sim$100 fs, remaining constant up to the picosecond timescale. The shift of $\omega_{0}$ is responsible for the plateau around 3 eV, observed in the $\Delta$$\mathcal{T}$/$\mathcal{T}$($\omega$,t) spectra (see Figs. \ref{fig3}a and \ref{fig3}b). The width of the Lorentzian profile is impulsively broadened to about 2 eV within the pulse duration and recovers a constant value of 1.2 eV. This feature is responsible for the broad variation of the transmissivity in the 2-3 eV range (see Fig. \ref{fig3}b). 

In Fig. \ref{fig4}b we report the imaginary part of $\epsilon_{L}$($\omega$) during the relaxation process when the 3.1 eV-pump is polarized along the b-axis. The resonance frequency appears at the equilibrium value $\omega_{0}$=3.67 eV and impulsively decreases to 3.1 eV during the pump pulse duration, recovering, approximately, the 3.67 eV-equilibrium value on the picosecond timescale.\cite{note3} The width of the Lorentzian profile, impulsively broadened to about 2 eV, recovers a constant value of 0.3 eV after the excitation. The combination of the lack of energy shift of the resonance frequency with the narrow variation of the lineshape explains the relatively small shape variations of the time-resolved spectra after $\sim$200 fs (see Fig. \ref{fig3}d), in comparison with the previous case.

Fig. \ref{fig4}c shows the dynamics of Im[$\epsilon_{L}$($\omega$)] when the system is excited by 1.55 eV pump pulses polarized parallel to the c-axis. Comparing the measured $\Delta$$\mathcal{T}$/$\mathcal{T}$($\omega$,t) to the results obtained at 3.1 eV-pump energy and with the same polarization (see Fig. \ref{fig4}a), the $\omega_{0}$-shift from 3.46 eV to 2.9 eV is similar, whereas the broadening of the linewidth is much more limited ($\gamma$$\sim$0.6 eV @ 1 ps). A clear oscillation of $\gamma$, with a period of $\sim$54 fs, is observed at 1.55 eV pump photon energy (see Fig. \ref{fig4}c). This oscillation can be attributed to the displacive coherent excitation \cite{Zeiger:1992} of an A$_g$ optical mode at 594 cm$^{-1}$, related to the bond-stretching vibration along the c-axis of apical oxygens.\cite{Popovic:1995} This finding is rather interesting since it confirms the strong coupling of the charge-transfer edge to the lattice structure.

Comparing the different set of data, the origin of the spectrally-resolved transmissivity variation in CuGeO$_3$, when delocalized excitons are photo-injected in the system, is clarified. A 0.6 eV shift of the resonance frequency of the charge-transfer edge is observed either when the pump photon energy is 3.14 eV or 1.55 eV. This result confirms that, when the charge transfer process is excited by c-polarized pump pulses, the holes, shared among the O-2$p$ orbitals and the Cu-3$d$ levels of the neighboring plaquettes, strongly interact with the Cu-3d equilibrium holes, creating a non-equilibrium population within the Cu-3$d$ bands.
For these reason, the 0.6 eV-shift can be attributed of the creation of a non-equilibrium distribution of excitations within the Cu-3$d$ bands. In particular, the measured resonance frequency decrease is compatible with a decrease of the energy needed to move a hole from the Cu-$3d$ levels to the O-$2p$ orbitals ($\delta \Delta_{pd} \sim$0.6 eV), since the higher crystal-field split $3d$ levels (see Fig. \ref{fig1}b) are occupied, on average, by the non-equilibrium population within the Cu-$3d$ bands, created by the pump pulse. 
On the contrary, the large broadening of the charge-transfer edge, measured when the system is excited by 3.14 eV c-polarized pump pulses, has no counterpart in the other configurations. This result suggests that the observed charge-transfer broadening is a genuine effect of the delocalized excitons photo-injected in the Cu$-$O$_4$ planes mainly altering the $T_{pd}$ term of the hamiltonian and leaving the system in a metastable non-thermal phase.\\

\begin{figure}
\includegraphics[keepaspectratio, bb= 60 50 920 310, clip,width=0.92\textwidth]{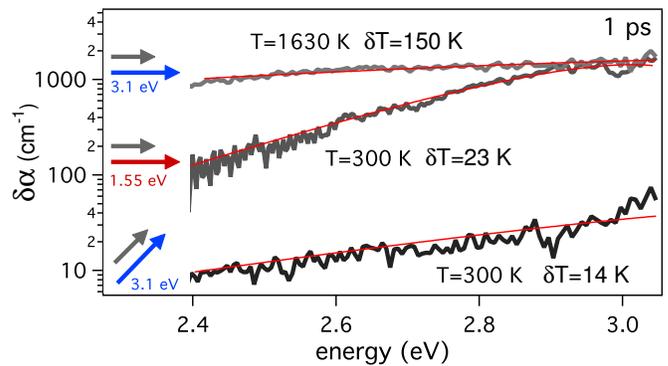}
\caption{\label{fig5} (Color online) Variation of the absorption coefficient, as a function of the probe energy, at 1 ps delay time. The solid lines are the fit of expression (\ref{diff_urbach}) to the data.}
\end{figure}

New insights into the origin of the metastable state created by the pump excitation are given by the analysis of the time-resolved spectra at 1 ps. At this time delay, the cooling processes related to electron-phonon scattering are over and, at t$>$1 ps, the spectra slowly decay with a several hundreds of ps timescale.
The non-thermal nature of the charge-transfer broadening, when delocalized excitons are photo-injected, can be argued recalling the Urbach's rule. Differentiating the expression (\ref{urbach}), $\delta \mathrm{T}$ being the differential increase of the temperature,
\begin{equation}
\label{diff_urbach}
\delta \alpha=\alpha_0e^{\left(\sigma\frac{E-E_0}{k\mathrm{T}}\right)}\left[\sigma\frac{E_0-E}{k\mathrm{T}^2}+\frac{\partial \sigma}{\partial \mathrm{T}}\frac{E-E_0}{k\mathrm{T}}\right]\delta \mathrm{T}
\end{equation}
and taking the logarithm of Eq. \ref{diff_urbach}, neglecting the $\partial \sigma/\partial \mathrm{T}$ term, we obtain:
\begin{equation}
\label{diff_urbach_log}
\mathrm{ln}(\delta \alpha)=\mathrm{ln}(\alpha_0\delta \mathrm{T})+\mathrm{ln}\left(\sigma\frac{E_0-E}{k\mathrm{T}^2}\right)+\sigma\frac{E-E_0}{k\mathrm{T}}
\end{equation} 
where the linear slope of ln($\delta \alpha$), typical of the Urbach's law, is corrected by a ln$\left(E_0-E\right)$ term. In Fig. \ref{fig5} we report, in log scale, the $\delta \alpha$, measured after 1 ps from the pump pulse, as a function of the probe energy. The non-linear correction to the standard Urbach's shape is evident, particularly in the c-polarized pump configurations. 

The expression (\ref{diff_urbach}) has been fitted to the data, being $\mathrm{T}$ and $\delta \mathrm{T}$ the only free parameters.\cite{note2} In the 1.55 eV$-$c-polarized and 3.14 eV$-$b-polarized configurations, the slopes of the measured $\delta \alpha$ are compatible with an effective local temperature of T=300 K , the temperature increase being $\delta \mathrm{T}$=23 K and $\delta \mathrm{T}$=14 K, respectively. These results must be compared with the estimated lattice heating due to the absorbed pump energy. This is given by: $\delta \mathrm{T}$=$I\alpha$/$C_{lat}$, $I$$\simeq$20 mJ/cm$^2$ being the absorbed pump fluence and $C_{lat}$=2.77 JK$^{-1}$cm$^{-3}$ the CuGeO$_3$ heat capacity.\cite{Weiden:1995} Assuming $\alpha$=500 cm$^{-1}$ and $\alpha$=250 cm$^{-1}$ in the 1.55 eV-c-polarized and 3.14 eV-b-polarized configurations,\cite{Pagliara:2002} we can estimate $\delta \mathrm{T}$=3.6 K and $\delta \mathrm{T}$=1.8 K, respectively. While the ratio of the estimated temperature increase (3.6 K/1.8 K$\simeq$2), related to the difference in the absorbed density of energy, is compatible with measured ratio (23 K/14 K$\simeq$1.6), the discrepancy of the absolute values can be attributed to the fact that the energy absorbed by the electronic transitions is delivered to a subset of optical lattice modes, i.e. strongly-coupled optical phonons (SCOPs).\cite{Kampfrath:2005} Therefore, the charge-transfer edge broadening, sensitive to the exciton scattering by optical phonons, reflects an effective temperature of the SCOPs higher than the temperature estimated considering the heat capacity of the whole lattice modes.

In contrast to the previous cases, when the system is excited by c-polarized 3.14 eV pump pulses and delocalized excitations are photoinjected in the O-$2p$ orbitals, the measured $\delta \alpha$ is fitted assuming an effective local temperature of 1630 K and a temperature increase of $\delta \mathrm{T}$=150 K (see Fig. \ref{fig5}), values incompatible with a quasi-thermal physical scenario. This result confirms that the linewidth broadening measured in this configuration (see Figs. \ref{fig3}b and \ref{fig4}a) is of non-thermal origin, i.e. it cannot be attributed to the local heating of the lattice.

\section{Conclusions}
\label{Sec6}
In conclusion, we have applied a time-and-frequency resolved technique to investigate the excited state dynamics of a model strongly-correlated system as CuGeO$_3$. The evolution of the dielectric function on the sub-ps timescale indicates that the impulsive injection of holes delocalized on the O-2$p$ oxygen orbitals strongly perturbs the potential experienced by electrons on the Cu-$3d$ levels, leaving the system in a non-thermal metastable (on the ps timescale) phase. 
The possibility to follow the decay dynamics of the electronic properties of a charge-transfer insulating cuprate system, impulsively driven in a non-equilibrium state, constitutes a further step toward the understanding of the puzzling physics of these systems. In addition, the experimental investigation of the non-thermal states obtained after the impulsive variation of the potential experienced by electrons in CuGeO$_3$ can be regarded as a benchmark test for the development of more realistic theoretical models for the non-equilibrium physics of strongly-correlated systems.\\

We acknowledge valuable discussions from D. Vollhardt, M. Kollar and M. Eckstein.

\bibliography{CuGeO3}

\begin{thebibliography}{31}
\bibitem[{not({\natexlab{a}})}]{note0}
\bibinfo{note}{The value $\tau$=10 fs has been extracted by fitting the
  convolution of the pump and probe pulses to a double exponential decay
  function to the experimental data.}
\bibitem[{not({\natexlab{c}})}]{note3}
\bibinfo{note}{Due to the small variations of the time-resolved spectra, the
  value of $\omega_0$=3.67 eV, determined through the differential fitting
  procedure, is subjected, on the sub-ps timescale, to an error larger than in
  the other configurations.}
\end{thebibliography}

\end{document}